\documentclass[12pt]{article}
\usepackage{amsfonts}
\usepackage{amsmath}
\usepackage{amsthm,amssymb}
\usepackage{algpseudocode}
\usepackage{algorithm}
\usepackage{graphicx,psfrag,epsf}
\usepackage{subcaption}
\usepackage{enumerate}
\usepackage{url}

\makeatletter
\renewcommand\@biblabel[1]{}
\makeatother

\newcommand{\blind}{1}
\input{tcilatex}
\begin{document}

\def\spacingset#1{\renewcommand{\baselinestretch}%
{#1}\small\normalsize} \spacingset{1}


\if1\blind
{
  \title{\bf Estimating matching affinity matrix under low-rank constraints}
  \author{Arnaud Dupuy\thanks{Dupuy gratefully acknowledges the support of a FNR grant C14/SC/8337045.}\hspace{.2cm}\\
CREA, University of Luxembourg, L-1511 Luxembourg \\
and \\
Alfred Galichon \\
Department of Economics and CIMS, NYU, New York, NY 10012\\
and \\
Yifei Sun\\
Department of Mathematics, CIMS, NYU, New York, NY 10012}
  \maketitle
} \fi

\if0\blind
{
  \bigskip
  \bigskip
  \bigskip
  \begin{center}
    {\LARGE\bf Estimating matching affinity matrix under low-rank constraints}
\end{center}
  \medskip
} \fi

\bigskip
\begin{abstract}
In this paper, we address the problem of estimating transport
surplus (a.k.a. matching affinity) in high dimensional optimal transport
problems. Classical optimal transport theory specifies the matching affinity
and determines the optimal joint distribution. In contrast, we study the
inverse problem of estimating matching affinity based on the observation of
the joint distribution, using an entropic regularization of the problem. To
accommodate high dimensionality of the data, we propose a novel method that
incorporates a nuclear norm regularization which effectively enforces a rank
constraint on the affinity matrix. The low-rank matrix estimated in this way
reveals the main factors which are relevant for matching.
\end{abstract}

\noindent%
\vfill

\newpage
\spacingset{1.45} 

\section{\protect\normalsize Introduction}

{\normalsize Optimal transport theory has attracted a lot of interest across
a number of scientific disciplines, from pure mathematics (Villani, 2003) to
various applications including machine learning (Benamou et al., 2015)
mathematical statistics (del Barrio et al., 1999) and economics (Galichon,
2016). The basic problem of optimal transport is how to form pairs of agents
drawn from two populations in order to maximize the total utility, also
called matching affinity. The resulting joint distribution of pairs is
called an optimal matching, also called optimal transport plan.  }

{\normalsize Most of the theory of optimal transport has focused on the
direct problem, namely solving for the optimal matching, taking the
matching affinities as given. In contrast, we consider in this paper the
\emph{inverse optimal transport problem}: given the observation of an
optimal matching, what is the affinity function for which this matching is
optimal\footnote{%
In the theoretical computer science literature, this problem is known as an
\emph{inverse assignment problem}, see Burkard et al. (2009), Section 6.7
and references therein.}? This problem arises naturally in the study of
two-sided matching markets, which appears in various fields of the
social sciences. In sociology and economics, one instance of these markets
is the \textquotedblleft marriage market,\textquotedblright\ following
Becker (1973)'s seminal analysis, where one observes the characteristics of
both partners in married couples (such as education, height, personality
traits, etc.), and one wants to infer (i)\ which characteristics attract or
repel each other the most, and (ii)\ what combinations of characteristics
are the most relevant for matching.  }

{\normalsize In models of matching markets, vectors of characteristics $x\in
\mathbb{R}^{d}$ for one side of the market and $y\in \mathbb{R}^{d^{\prime
}} $ for the the other side are available, and the joint distribution $\hat{%
\pi}\left( x,y\right) $ across matched pairs is observed, and we are
interested in estimating the matching affinity function $\Phi \left(
x,y\right) $. Broadly speaking, models of matching markets are divided into
three categories: scalar index models, discrete models, and multivariate
models, which we will now briefly survey.  }

{\normalsize \emph{Scalar index models}. A number of papers use scalar index
models: they assume that agents match on a pair of scalar indices $\tilde{x}%
=u^{\top }x$ and $\tilde{y}=v ^{\top }y$, which are weighted sums of
partners' characteristics. Following a suggestion by Becker (1973), a number
of papers have used canonical correlation or linear regression techniques in
order to estimate the weight vectors $u$ and $v$; see for instance Suen and
Lui (1999), Gautier et al. (2005), Lam and Schoeni (1993, 1994), and Jepsen
(2005), and a caution against the misuse of these techniques in Dupuy and
Galichon (2015). A more robust ways to estimate the weight vectors has been
suggested by Tervi\"{o} (2003) using rank correlation. See also Chiappori et
al. (2012).  }

{\normalsize \emph{Discrete models}. Following a seminal paper by Choo and
Siow (2006), a number of recent papers (Fox, 2010; Chiappori et al., 2016;
Galichon and Salani\'{e}, 2015) have assumed that agents match based on
discrete characteristics, either categorical variables like ethnicity, or
binned, such as the income bracket. However, the binning of cardinal
variables may be problematic as the results may depend heavily on the
arbitrary choice of the thresholds. Therefore, these models suffer from
limitations when dealing with non-categorical variables.  }

{\normalsize \emph{Continuous models}. More recently, a continuous model has
been proposed by Dupuy and Galichon (2014), where the matching affinity is
bilinear with respect to the matched pairs' characteristics, i.e. is given
by $x^{\top }Ay$, where $A$, called the \emph{affinity matrix} is a $d\times
d^{\prime }$ matrix is to be estimated. This model enables weighted
interactions between any pair of characteristics. Of course, when the rank
of $A$ is one, $A=\lambda uv^{\top }$, and one recovers the scalar index
models discussed above. But as soon as the rank of $A$ is greater than one,
a pair of scalar indices on each side of the market would not be sufficient to describe the matching
affinity. Dupuy and Galichon (2014) propose a moment matching procedure to
estimate $A$, which can be computed via convex optimization. However, as
soon as the number of characteristics goes large, the number of parameters
to be estimated grows quadratically, potentially leading to an overfit. }

{\normalsize In this paper, we propose a novel method for solving the
inverse optimal transport problem in a high-dimensional setting, where we
estimate the affinity matrix $A$ under a rank constraint in order to capture
the relevant dimensions of interaction on which matching occurs. Two
applications to the marriage market are proposed that each highlight
different features of the proposed method. The first application uses the
same data as in Dupuy and Galichon (2014) and illustrates how our method
allows one to identify the impact of narrowly defined personality traits
without having to aggregate these into aggregate traits prior to the estimation
of the affinity matrix as in that paper. The second
application is performed on data compiled in Banerjee et al. (2013) about
the role of castes in the Indian marriage market and illustrates the
usefulness of our method when one is dealing with categorical or ordinal
variables and one does not want to either ex ante aggregate categories or
assume some cardinal scale prior to estimating the affinity matrix.  }

{\normalsize The rest of the paper is organized as follows. Section 2
presents the matching equilibrium model and introduces the concept of
affinity matrix. Section 3 describes the maximum likelihood estimation of
the affinity matrix, including a low-rank regularized version. Section 4
present applications to two marriage markets datasets. Section 5 concludes
the paper.  }

\section{\protect\normalsize The model}

{\normalsize We first briefly recall the optimal transport problem; see
Villani (2003 and 2008) for more. Given two probability distributions $\mu
_{1}$ and $\mu _{2}$ over $\mathbb{R}^{d}\times \mathbb{R}^{d^{\prime }}$,
the optimal transport problem is defined as%
\begin{equation}
\max_{\pi \in \Pi \left( \mu _{1},\mu _{2}\right) }\mathbb{E}_{\pi }\left[
\Phi \left( X,Y\right) \right]  \label{Wasserstein}
\end{equation}%
where $\Phi \left( x,y\right) $\ is the measure of affinity between two
agents $x\in \mathbb{R}^{d}$ and $y\in \mathbb{R}^{d^{\prime }}$ on each side of the market, and $%
\Pi \left( \mu _{1},\mu _{2}\right) $ is the set of distributions $\pi
\left( x,y\right) $ with marginal distributions $\mu _{1}$ and $\mu _{2}$.
Problem~(\ref{Wasserstein}) is the\emph{\ Monge-Kantorovich problem} of
optimal transport.  }

\subsection{\protect\normalsize Optimal solution vs equilibrium}

{\normalsize The optimization problem~(\ref{Wasserstein}) yields a
centralized solution where a central planner would decide which pairs to
form. However, most matching markets (including the marriage market which we
study in this paper) are \emph{decentralized markets}, in which agents
decide based on their own interest, leading to an equilibrium. It follows
from the work of Becker (1973) and Shapley and Shubik (1971) that the
centralized and the decentralized problems are equivalent. We sketch the
argument as follows.  }

{\normalsize In decentralized problems, an \emph{outcome} is the
specification of a matching $\pi \in \Pi \left( \mu _{1},\mu _{2}\right) $,
and of individual payoffs $u\left( x\right) $ and $v\left( y\right) $, which
are attained by agents of respective types $x$ and $y$. The outcome is
called \emph{stable} when%
\begin{equation}
u\left( x\right) +v\left( y\right) \geq \Phi \left( x,y\right) ~\forall x,y.
\label{dualIneq}
\end{equation}%
Stability is a required condition for equilibrium. Indeed, if~(\ref{dualIneq}%
) were not to hold, then $\epsilon =\Phi \left( x,y\right) -u\left( x\right)
-v\left( y\right) $ would be strictly positive, and thus by matching
together, $x$ and $y$ could attain $u\left( x\right) +\epsilon /2$ and $%
v\left( y\right) +\epsilon /2$, which is strictly more than their
equilibrium payoffs $u\left( x\right) $ and $v\left( y\right) $. At the same
time if $x$ and $y$ are matched at equilibrium under $\pi ^{eq}$, then
feasibility imposes that $u\left( x\right) +v\left( y\right) =\Phi \left(
x,y\right) $. Thus, taking expectations of both sides with respect to $\pi
^{eq}$ will get
\begin{equation}
\mathbb{E}_{\pi ^{eq}}\left[ \Phi \left( X,Y\right) \right] =\mathbb{E}_{\pi
^{eq}}\left[ u\left( X\right) +v\left( Y\right) \right] =\mathbb{E}_{\mu
_{1}}\left[ u\left( X\right) \right] +\mathbb{E}_{\mu _{2}}\left[ v\left(
Y\right) \right] .  \label{feasibilityCond}
\end{equation}
}

{\normalsize Hence, $\pi ^{eq}\in \Pi \left( \mu
_{1},\mu _{2}\right) $  is defined as an equilibrium matching
whenever there exists functions $u$ and $v$v such that both conditions (\ref{dualIneq}) and (\ref{feasibilityCond}) hold.
}

{\normalsize Let us now show that if $\pi ^{eq}$ is an equilibrium, then it
is a solution of~(\ref{Wasserstein}). Consider $\pi ^{opt}$ a solution of
problem~(\ref{Wasserstein}). Taking expectations of both sides of~(\ref%
{dualIneq}) with respect to $\pi ^{opt}$ gets
\begin{equation*}
\mathbb{E}_{\pi ^{opt}}\left[ \Phi \left( X,Y\right) \right] \leq \mathbb{E}%
_{\pi ^{opt}}\left[ u\left( X\right) +v\left( Y\right) \right] =\mathbb{E}%
_{\mu _{1}}\left[ u\left( X\right) \right] +\mathbb{E}_{\mu _{2}}\left[
v\left( Y\right) \right]
\end{equation*}%
where the latter equality comes from the fact that $\pi ^{opt}\in \Pi \left(
\mu _{1},\mu _{2}\right) $. Hence, $\mathbb{E}_{\pi ^{opt}}\left[ \Phi
\left( X,Y\right) \right] \leq \mathbb{E}_{\pi ^{eq}}\left[ \Phi \left(
X,Y\right) \right] $, but by definition of $\pi ^{opt}$, these two
quantities coincide and $\pi ^{eq}$ is optimal for the centralized problem~(%
\ref{Wasserstein}). Hence, the decentralized solution (equilibrium
matching)\ coincides with the centralized solution~(optimal matching).  }

{\normalsize However, the analysis above assumes that the
existence of a matching between two partners is purely deterministic
given partners' observed characteristics, which is not
realistic. In order to
allow for some randomness arising from agent's unobserved
heterogeneity in the matching process, we shall make use of a regularized version of the
optimization formulation~(\ref{Wasserstein}) in order to perform the estimation
of $\Phi $.  }

\subsection{\protect\normalsize Modeling heterogeneity}

{\normalsize It is a well-known result in optimal transport theory (see
Villani, 2008, Chapter 9) that, under suitable assumptions on $\Phi $, the
optimal matching will be \emph{pure}, in the sense that any $x$ is
matched deterministically to a unique $y=T\left( x\right) $ for some
bijective map $T $; in other words, the conditional distribution $\pi \left(
y|x\right) $ of $y$ given $x$, is reduced to a single point mass. Clearly,
in the presence of unobserved heterogeneity, this is no longer the case. Our
approach to modeling uncertainty consists in adding an entropic
regularization term in~(\ref{Wasserstein}), leading to%
\begin{equation}
\max_{\pi \in \Pi \left( \mu _{1},\mu _{2}\right) }\mathbb{E}_{\pi }\left[
\Phi \left( X,Y\right) -\sigma \ln \pi \left( X,Y\right) \right]
\label{RegulWasserstein}
\end{equation}%
where $\sigma >0$ is a temperature parameter, so that setting $\sigma =0$
recovers program~(\ref{Wasserstein}).  }

{\normalsize Recently a number of authors have studied such a regularized
version of the Monge-Kantorovich problem (see for instance Benamou et al.,
2015; Galichon and Salani\'{e}, 2015 and references therein). One notable
feature of~(\ref{RegulWasserstein}) is that the optimal matching $\pi
\left( x,y\right) $ has form
\begin{equation*}
\pi \left( x,y\right) =a\left( x\right) b\left( y\right) \exp \left( \Phi
\left( x,y\right) /\sigma \right) ,
\end{equation*}%
where $a\left( x\right) $ and $b\left( y\right) $ are set by imposing the
constraint $\pi \in \Pi \left( \mu _{1},\mu _{2}\right) $, that is%
\begin{equation*}
\left\{
\begin{array}{c}
\int a\left( x\right) b\left( y\right) \exp \left( \Phi \left( x,y\right)
/\sigma \right) dy=\mu _{1}\left( x\right) \\
\int a\left( x\right) b\left( y\right) \exp \left( \Phi \left( x,y\right)
/\sigma \right) dx=\mu _{2}\left( y\right)%
\end{array}%
\right. .
\end{equation*}%
As a result, $a\left( x\right) $ and $b\left( y\right) $ can be obtained by
the iterated proportional fitting procedure (IPFP), a.k.a. Sinkhorn's algorithm,
which is presented in algorithm \ref{sinkhorn}.
\begin{algorithm}
\caption{IPFP}\label{sinkhorn}
\begin{algorithmic}
\Require $b(y),\mu_1(x), \mu_2(y), \Phi(x,y), \sigma$
\While {not converged}
	\State $a\left( x\right)  \gets \mu _{1}\left( x\right)/ \int b\left(y\right)\exp \left( \Phi \left( x,y\right) /\sigma \right) dy$
	\State $b\left( y\right)  \gets \mu _{2}\left( y\right) / \int a\left(x\right) \exp \left( \Phi \left( x,y\right) /\sigma \right) dx$
\EndWhile
\Ensure $a(x), b(y)$
\end{algorithmic}
\end{algorithm}


\subsection{\protect\normalsize Parameterization of the affinity function}

{\normalsize We assume the simple parameterization of $\Phi $ as a bilinear
form associated to some \emph{affinity matrix} $A$, namely
\begin{equation}
\Phi _{A}\left( x,y\right) =x^{\top }Ay.  \label{parameterization}
\end{equation}%
This functional form will capture the interaction effects between the
various dimensions of the characteristics. The sign of $A_{ij}$ indicates
that there is attractive (if positive $A_{ij}>0$)\ or repulsive (if $%
A_{ij}<0 $) energy between dimension $i$ of $x$ and dimension $j$ of $y$. On
the contrary, $A_{ij}=0$ means that there is no interaction between $x_{i}$
and $y_{j}$.  }

{\normalsize By positive homogeneity, we can normalize the temperature
parameter $\sigma $ in front of the entropic term to $\sigma =1$. Indeed,
the solution of the problem with affinity function $\Phi $\ and temperature $%
\sigma $ coincides with the solution of the problem with affinity function $%
\Phi /\sigma $\ and temperature one. Hence, we define
\begin{equation}
\mathcal{W}\left( A\right) =\max_{\pi \in \Pi \left( \mu _{1},\mu
_{2}\right) }\mathbb{E}_{\pi }\left[ \Phi _{A}\left( X,Y\right) -\ln \pi
\left( X,Y\right) \right] ,  \label{defWA}
\end{equation}
}

{\normalsize As before, the optimal matching $\pi ^{A}$ retains the
form
\begin{equation}
\pi ^{A}\left( x,y\right) =a\left( x\right) b\left( y\right) \exp \left(
\Phi _{A}\left( x,y\right) \right) ,  \label{sinkhornBis}
\end{equation}%
where $a\left( x\right) $ and $b\left( y\right) $ are computed by the IPFP
algorithm \ref{sinkhorn}. It follows directly from expression~(\ref%
{sinkhornBis})\ that%
\begin{equation*}
\frac{\partial ^{2}\ln \pi ^{A}\left( x,y\right) }{\partial x_{i}\partial
y_{j}}=A_{ij}
\end{equation*}%
which provides a nice interpretation of $A$ as the matrix of
cross-derivatives of the log-likelihood of a matched $\left( x,y\right) $
pair. In the sequel, we shall focus on the estimation of the affinity matrix
$A$.  }

\section{\protect\normalsize Maximum likelihood estimation of the affinity
matrix}

{\normalsize We would like to estimate $A$\ based on an i.i.d. sample of
matched pairs $\left( x^{(k)},y^{(k)}\right) $, $k=1,...,N$, where $x^{(k)}$
and $y^{(k)}$ are respectively $d$ and $d^{\prime }$-dimensional vectors of
characteristics, and the observed matching is defined as%
\begin{equation*}
\hat{\pi}\left( x,y\right) =\frac{1}{N}\sum_{k=1}^{N}\delta \left(
x-x^{(k)}\right) \delta \left( y-y^{(k)}\right) .
\end{equation*}
}
\subsection{\protect\normalsize Unconstrained maximum likelihood}

{\normalsize As implied by the next result, the likelihood function turns
out to have a particularly tractable form and is globally concave.  }

{\normalsize
\begin{proposition}
{\normalsize \label{prop:logLikelihood} (a) The log-likelihood $l\left( A;%
\hat{\pi}\right) $ of observation $\hat{\pi}$ at parameter value $A$ is
given by%
\begin{equation}
l\left( A;\hat{\pi}\right) =N\mathbb{E}_{\hat{\pi}}\left[ \log \pi
^{A}\left( X,Y\right) \right] =N\left\{ \mathbb{E}_{\hat{\pi}}\left[ \Phi
_{A}\left( X,Y\right) \right] -\mathcal{W}\left( A\right) \right\} .
\label{logLikelihood}
\end{equation}%
(b) It is a concave function of }$A$, and its gradient is given by
\begin{equation}
\nabla l\left( A;\hat{\pi}\right) =N\left\{ \mathbb{E}_{\hat{\pi}}\left[ X_{i}Y_{j}
\right] -\mathbb{E}_{\pi ^{A}}\left[ X_{i}Y_{j}\right] \right\} .
\label{focMLE}
\end{equation}
\end{proposition}
}

{\normalsize
\begin{proof}
(a) {\normalsize The log-likelihood of a pair $\left( x^{(k)},y^{(k)}\right)
$ is given by $\log \pi ^{A}\left( x^{(k)},y^{(k)}\right) $. As the pairs
are independently sampled, the log-likelihood of the matching $\hat{\pi}$ is
given by $l\left( A;\hat{\pi}\right) =\sum_{k=1}^{N}\log \pi ^{A}\left(
x^{(k)},y^{(k)}\right) =N\mathbb{E}_{\hat{\pi}}\left[ \log \pi ^{A}\left(
X,Y\right) \right] $. It follows from~(\ref{sinkhornBis}) that~}$\mathbb{E}_{%
\hat{\pi}}\left[ \log \pi ^{A}\left( X,Y\right) \right] =\mathbb{E}_{\hat{\pi%
}}\left[ \Phi _{A}\left( X,Y\right) \right] +\mathbb{E}_{\hat{\pi}}\left[
\log a\left( X\right) +\log b\left( Y\right) \right] $, but as $\pi ^{A}$
and $\hat{\pi}$ both belong to $\Pi \left( \mu _{1},\mu _{2}\right) $, it
follows that
\begin{eqnarray*}
&&\mathbb{E}_{\hat{\pi}}\left[ \log \pi ^{A}\left( X,Y\right) \right] \\
&=&\mathbb{E}_{\hat{\pi}}\left[ \Phi _{A}\left( X,Y\right) \right] -\mathbb{E%
}_{\pi ^{A}}\left[ \Phi _{A}\left( X,Y\right) \right] +\mathbb{E}_{\pi ^{A}}%
\left[ \Phi _{A}\left( X,Y\right) \right] +\mathbb{E}_{\pi ^{A}}\left[ \log
a\left( X\right) +\log b\left( Y\right) \right] \\
&=&\mathbb{E}_{\hat{\pi}}\left[ \Phi _{A}\left( X,Y\right) \right] -\mathbb{E%
}_{\pi ^{A}}\left[ \Phi _{A}\left( X,Y\right) \right] +\mathbb{E}_{\pi ^{A}}%
\left[ \log \pi ^{A}\left( X,Y\right) \right] \\
&=&\mathbb{E}_{\hat{\pi}}\left[ \Phi _{A}\left( X,Y\right) \right] -\mathcal{%
W}\left( A\right),
\end{eqnarray*}
hence $l\left( A;\hat{\pi}\right) =N\left\{ \mathbb{E}_{%
\hat{\pi}}\left[ \Phi _{A}\left( X,Y\right) \right] {\normalsize -}\mathcal{W%
}\left( A\right) \right\} $.

(b) $\mathbb{E}_{\hat{\pi}}\left[ \Phi _{A}\left( X,Y\right) \right] $ is
linear in $A$, and $\mathcal{W}\left( A\right) $ is convex in $A$, so
{\normalsize $l\left( A;\hat{\pi}\right) $ is concave. By the envelope
theorem, }$\nabla l\left( A;\hat{\pi}\right) =N\left\{ \mathbb{E}_{\hat{\pi}}%
\left[X_{i}Y_{j}\right] -\mathbb{E}_{\pi ^{A}}\left[ X_{i}Y_{j}\right] \right\} $.
\end{proof}
}

{\normalsize Thus, conditions$~$(\ref{focMLE}) imply that the maximum
likelihood estimator$\ \hat{A}$ should solve%
\begin{equation}
\mathbb{E}_{\pi ^{A}}\left[ X_{i}Y_{j}\right] =\mathbb{E}_{\hat{\pi}}\left[
X_{i}Y_{j}\right]   \label{momentMatching}
\end{equation}%
for every pair $i\in \{1,\cdots ,d\}$ and $j\in \{1,\cdots ,d^{\prime }\}$,
which thus turns out to be equivalent to the moment matching procedure of
Dupuy and Galichon (2014). Hence, assuming w.l.o.g. that $X$ and $Y$ are
centered at 0, this implies that $\hat{A}$ is the value of the parameter
such that the predicted covariance matrix $cov_{\pi ^{A}}\left( X,Y\right) $
will match the observed one $cov_{\hat{\pi}}\left( X,Y\right) $. }

{\normalsize One important advantage of the concavity of the log-likelihood
function $l\left( A;\hat{\pi}\right) $ is that various additional
regularizations can be incorporated into the estimation procedure. One could
constrain $A$ to be entry-wise nonnegative so that only attractive
interactions are considered. One could also assume $A$ is sparse, so that
only a small number of pairs of characteristics interact. In this paper, we
are concerned with the case when only a small number of dimensions, which
are linear combinations of the characteristics, interact. One shall then
need to impose a requirement that the rank of $A$ is small. The next
sections propose an effective method for doing so which is implemented on
two marriage market datasets.  }

\subsection{\protect\normalsize Low-rank regularization}

{\normalsize In some situations, two scalar dimensions $\tilde{x}$ and $%
\tilde{y}$, obtained linearly from $x$ and $y$ via $\tilde{x}=u^{\top }x$
and $\tilde{y}=v^{\top }y$, suffice to explain the solution $\hat{\pi}$,
where $u$ and $v$ are two unit vectors of weights. In this case, $A$ is
simply a scalar multiple of rank one matrix $uv^{\top }$. More generally,
when the rank of $A$ is equal to $r$, the singular value decomposition (SVD)
of $A$ yields
\begin{equation}
A=USV^{\top },  \label{SVD}
\end{equation}%
where $S$ is a diagonal $r\times r$ matrix with strictly positive diagonal
entries (called singular values) in the decreasing order, and $U$ and $V$
are two semi-orthogonal $d\times r$ matrices. In this case, the total
interaction term is $x^{\top }Ay=\tilde{x}^{\top }S\tilde{y}$, where $\tilde{%
x}=U^{\top }x$ and $\tilde{y}=V^{\top }y$ are the relevant dimensions of
interaction. Note that $x^{\top }Ay$ requires to sum over $d\times d^{\prime
}$ interaction terms whereas $\tilde{x}^{\top }S\tilde{y}$ only requires to
sum over $r\leq \min \left\{ d,d^{\prime }\right\} $ interaction terms.
Moreover, each singular value can be interpreted as the weight of the
interaction between the corresponding relevant dimensions of $\tilde{x}$ and
$\tilde{y}$ in the total interaction term.  }

{\normalsize One can incorporate the rank constraint into the maximization
of the likelihood, whose expression is given in proposition~\ref%
{prop:logLikelihood}, yielding%
\begin{eqnarray*}
\max_{A} &&l\left( A;\hat{\pi}\right) \\
s.t.~ &&rk\left( A\right) \leq r.
\end{eqnarray*}%
However, the general rank-constrained problem is non-convex and NP-hard, see
Fazel (2002). A natural convex relaxation of the problem is done by
replacing the rank of $A$ by its nuclear norm (see e.g. Fazel, 2002; Recht
et al., 2010), $\Vert A\Vert _{\ast }$, defined as the sum of the singular
values of $A$. This yields a modified formulation of the problem as
\begin{equation}
\min_{A}\left\{ \mathcal{W}\left( A\right) -\mathbb{E}_{\hat{\pi}}\left[
\Phi _{A}\left( X,Y\right) \right] +\lambda \left\Vert A\right\Vert _{\ast
}\right\}  \label{minKLnuclear}
\end{equation}%
where $\lambda \geq 0$ is the Lagrange multiplier of the nuclear norm
constraint. Measuring the complexity of the model by the rank of the
affinity matrix, equation (\ref{minKLnuclear}) indicates that for $\lambda
=0 $, one accepts the full complexity of the model and perform exact
likelihood maximization whereas, for large values of $\lambda $, one
simplifies the model and deviates from exact likelihood maximization. Hence,
the parameter $\lambda $ can be thought of as a parameter controlling the
trade-off between exact likelihood maximization and the complexity of the
model.  }

{\normalsize The computation for problems involving the nuclear norm can be
efficiently carried out using the proximal gradient descent method with
guaranteed convergence (see e.g. Toh and Yun, 2010). As noted in the
previous section, $l\left( A;\hat{\pi}\right) $\ is continuously
differentiable with respect to $A$, and its gradient is given in expression (%
\ref{focMLE}). We now describe our complete procedure in algorithm \ref%
{proxAlgo}.  }

{\normalsize
\begin{algorithm}
\caption{Proximal gradient descent algorithm for problem (\ref{minKLnuclear})}\label{proxAlgo}
\begin{algorithmic}
\Require $A$, step size $t$, matched pairs $\left( x^{(k)},y^{(k)}\right) $%
, $k=1,...,N$
\While {not converged}
	\State Using the IPFP algorithm \ref{sinkhorn} to compute the optimal matching $\pi^A$
	\State$A\gets A-t\left(\sum_{i,j=1}^{N}(\pi^A_{ij}-\hat{\pi}_{ij})x^{(i)}(y^{(j)})^{\top}\right)$
	\State $[U, \mathsf{diag}(s_1,\cdots,s_d), V] =  \mathsf{SVD}(A)$
	\State$A\gets U\mathsf{diag}((s_1-t\lambda)_+,\cdots,(s_d-t\lambda)_+)V^\top $
\EndWhile
\Ensure $A$
\end{algorithmic}
\end{algorithm}
}

{\normalsize Additionally, we note that the nuclear norm regularization
prevents overfitting the covariance mismatch $\|\mathbb{E}_{\pi ^{A}}\left[
XY^{\top }\right] -\mathbb{E}_{\hat{\pi}}\left[ XY^{\top }\right]\|_F$,
where $\|\cdot\|_F$ is the Frobenius norm of a matrix and which one recalls
from expression (\ref{momentMatching}) will be exactly equal to 0 without
the nuclear norm regularization. Indeed, given $U$ and $V$ defined in
expression~(\ref{SVD}), the first order optimality conditions (Watson, 1992)
are
\begin{equation}
\mathbb{E}_{\pi ^{A}}\left[ XY^{\top }\right] -\mathbb{E}_{\hat{\pi}}\left[
XY^{\top }\right] +\lambda UV^{\top }+N=0,  \label{foc}
\end{equation}%
where $N$ satisfies $U^{\top }N=0$, $NV=0$, and $\Vert N\Vert _{2}\leq
\lambda $, with $\left\Vert N\right\Vert _{2}$ being the \textit{spectral
norm} of $N$. Equation (\ref{foc}) indicates that $\mathbb{E}_{\pi ^{A}}%
\left[ XY^{\top }\right] -\mathbb{E}_{\hat{\pi}}\left[ XY^{\top }\right] $
and $A$ have simultaneous SVDs. Moreover, the singular values of $\mathbb{E}%
_{\pi ^{A}}\left[ XY^{\top }\right] -\mathbb{E}_{\hat{\pi}}\left[ XY^{\top }%
\right]$ corresponding to the strictly positive singular values of $A$ will
be exactly equal to $\lambda $, while the ones corresponding to the zero
singular values of $A$ will be less than or equal to $\lambda $. Thus, by
varying $\lambda$, the covariance mismatch, which equals the $l_{2}$-norm of
the singular values of $\mathbb{E}_{\pi ^{A}}\left[ XY^{\top }\right] -%
\mathbb{E}_{\hat{\pi}}\left[ XY^{\top }\right]$, will change as well.  }

{\normalsize We select the best $\lambda$ by repeating a five-fold
cross-validation (CV) twice, resulting in ten different experiments. In each
of the CV procedure, the whole dataset is randomly split into five parts
with equal size. For each $\lambda$, we estimate $A$ via (\ref{minKLnuclear}%
) using 4 parts and record both $\mathcal{W}\left( A\right) -\mathbb{E}_{%
\hat{\pi}}\left[ \Phi _{A}\left( X,Y\right) \right] $ and $\Vert \mathbb{E}%
_{\pi ^{A}}\left[ XY^{\top }\right] -\mathbb{E}_{\hat{\pi}}\left[ XY^{\top }%
\right] \Vert _{F}$ evaluated on the remaining part. From this we obtain an
estimated prediction error curve as a function of $\lambda$, and we select
the $\lambda$ value that minimizes both errors.  }

\section{\protect\normalsize Application to marriage market data}

{\normalsize We apply the low-rank optimal transport method to the case of
bipartite matching in the marriage market. We choose two applications, each
corresponding to a data set with features allowing us to test different
aspects of our method. The first application revisits the data set used in
Dupuy and Galichon (2014). This data confronts the analyst with the problem
of selecting from a large set of observed characteristics of spouses, those
that are important for matching affinities. The second application uses data
compiled in Banerjee et al. (2013) for which the analyst has access to
categorical variables describing spouses' observable characteristics. To
take into account the effect of each categorical variable on matching
affinities, the analyst needs to create as many dummy variables as
categories distinguished, hence increasing rapidly the dimensionality of the
affinity matrix.  }

{\normalsize In both these applications, the analyst faces the difficult
task of estimating an affinity matrix whose size is large, being the product
of the number of observed characteristics of spouses, relative to the number
of observations. The high ratio of parameters to observations creates
overfitting concerns. A solution would be to construct combinations of the
observed characteristics prior to the estimation, hence reducing the number
of parameters of the associated affinity matrix. However, the construction
of these combinations of characteristics requires the analyst to define
weights based on prior information about matching affinity. In contrast, our
low-rank optimal transport method allows the analyst to simultaneously
estimate the affinity matrix while selecting the relevant combinations of
characteristics using weights derived from the information contained in the
affinity matrix itself.  }

\subsection{\protect\normalsize Personality traits}

{\normalsize Our first application uses the Dutch Household Survey (DHS) ran
by the Dutch National Bank. In particular, a representative sample of 1,155
young couples observed in the period 1993-2002 in the Netherlands was
constructed following the procedure outlined in Dupuy and Galichon (2014).
In this sample, the analyst has access to detailed information about
spouses' characteristics such as education, height, Body Mass Index (BMI)%
\footnote{{\normalsize Weight in Kg divided by the square of height in
meters.}} and subjective health, but also about personality traits and
attitude towards risk. Personality traits are herewith recovered by
administrating the 16 Personality Factors test (16PF test) to respondents
(see e.g. Cattell et al., 1993). This test consists in a 16-item questionnaire
where each item corresponds to a primary factor describing a facet of one's
personality. Attitude towards risk is recovered using a similar approach
(see e.g. Donkers and Van Soest, 1999). A 6-item questionnaire about risk
attitude is administrated to the respondents, each item corresponding to a
primary factor describing a facet of one's attitude towards risks.  }

{\normalsize In this application, the objective is to estimate matching
affinities from the sample of 1,155 couples with characteristics $(X,Y)$,
where $X$ and $Y$ contain each $26$ variables: education, height, BMI,
subjective health and the 16 primary factors of personality traits and 6
primary factors of risk attitude. The associated affinity matrix has $%
26\times 26=676$ parameters to be estimated, hence a ratio of 0.58
parameters per observation. Dupuy and Galichon (2014) substantially reduced
the dimensionality of the model by constructing 5 global factors of
personality traits and 1 global factor of attitude towards risk. They relied
on the psychology literature that shows that 5 global factors, often
referred to as the \textquotedblleft big 5," providing an overview of one's
personality can be derived from the primary factors of the 16PF using
methods such as Factor Analysis. These 5 global factors are (orthogonal)
linear combinations of the 16 primary factors. Similarly, as is standard in
the economic literature (see e.g. Donkers and Van Soest, 1999), a single
global factor providing an overview of attitude towards risk can be derived
as a linear combination of the underlying 6 primary factors. As a result,
Dupuy and Galichon (2014) were able to estimate a reduced affinity matrix of
dimension $10\times 10=100$, with a ratio of $0.09 $ parameters per
observation. However, this requires to assume that i) either all or none of
the primary factors belonging to a global factor matter and ii) their
relative importance is proportional to their relative weight in the global
factor. There are no reasons to expect this should hold universally since
the weights used to create the global factors are chosen so as to provide an
overview\ of an individual's personality or attitude towards risk and not to
capture matching affinities. In contrast, our low-rank optimal transport
approach, allows us to estimate the affinity matrix of size $676$ associated
with the primary factors while creating the relevant combinations of these
factors that matter for matching affinities.  }

{\normalsize We use the low-rank optimal transport approach to estimate the
affinity matrix $A$ when considering characteristics including the primary
factors. Inspection of Figure \ref{DutchCV} indicates that $\lambda =0.15$
gives slightly lower values of the CV errors of both $\mathcal{W}\left(
A\right) -\mathbb{E}_{\hat{\pi}}\left[ \Phi _{A}\left( X,Y\right) \right] $
and $\Vert \mathbb{E}_{\pi ^{A}}\left[ XY^{\top }\right] -\mathbb{E}_{\hat{%
\pi}}\left[ XY^{\top }\right] \Vert _{F}$ than $\lambda =0.1$ does. Since $%
\lambda =0.15$ achieves this result with a lower rank of $A$, we use this
value as the coefficient for the nuclear norm regularization. The left panel
of Figure \ref{DutchSingular} reveals the rank of the affinity matrix is 12
hence indicating that only 12 relevant dimensions matter for matching
affinities. Of those 12 relevant dimensions, the first three alone explain
about 50\% of the total matching affinity as indicated in the right panel of
Figure \ref{DutchSingular}.  }

{\normalsize The loadings of the first three dimensions, reported in Table %
\ref{DutchTable}, reveal several important results. First, as in Dupuy and
Galichon (2014), we do find that the first relevant dimension loads
principally on education, i.e. 0.85 for men and 0.83 for women respectively,
whereas the second and third dimensions load principally on personality
traits and attitude towards risk. However, using the primary factors rather
than the global factors as in Dupuy and Galichon (2014), we find that
although conscientiousness matters for both men and women, the underlying
primary factors at play differ across gender. For women, the primary factor
\textquotedblleft easily hurt, offended,\textquotedblright\ belonging to the
global factor conscientiousness, is the most important characteristic in the
second dimension with a loading of 0.71. For men, the primary factor
\textquotedblleft easily hurt, offended\textquotedblright\ plays also an
important role (loading of magnitude 0.42), but the primary factor
\textquotedblleft disciplined,\textquotedblright\ also belonging to the
global factor conscientiousness, is the most important one with a loading of
0.52. These results clearly illustrates that although conscientiousness
matters, not all of its primary constituents do and different aspects matter
differently for men and women.  }

{\normalsize A similar type of results holds for the third dimension, which
loads on some but not all of the items measuring attitude towards risk.
However, this dimension also loads on other variables such as height, BMI
and subjective health, making its interpretation more difficult.  }

\subsection{\protect\normalsize Castes in India}

{\normalsize The second application of our method is on data compiled by
Banerjee et al. (2013). These data were collected based on interviews of
Hindus families that placed a matrimonial ad in the major Bengali newspaper
in India, between October 2002 and March 2003\footnote{%
In India, marriages are often a family affair, with parents or relatives of
prospective brides or grooms placing a matrimonial ad in a newspaper.}. A
year after, 289 brides and grooms that had gotten married or engaged were
interviewed a second time, resulting in, as labeled in Banerjee et al.
(2013), the \textquotedblleft matches\textquotedblright\ data. We use the
sample of 284 couples for which information about the caste of both spouses
is available. Based on this information, we create 8 dummy variables, one
for each of the 8 main Hindus castes in India. In addition, the data
contains information about the height, education, family origin (from west
Bengal or not), the number of older (younger) brothers, the number of older
(younger) sisters, income and per capita consumption of each spouse. To
avoid deleting too many observations because of missing information on
height and per capita consumption we proceeded as follows. For each of these
variables, we replaced missing values by the sample mean and created a dummy
variable indicating missing. Both, the imputed variable and the missing
indicator were included into the set of characteristics considered in the
analysis. Our working data set contains a sample of 284 couples with 19
characteristics observed for each spouse.  }

{\normalsize In this application, the analyst is therefore confronted with
the problem of estimating an affinity matrix $A$\ of size $19\times 19=361$
using a sample of $284$ couples. The ratio of 1.27 parameters per
observation\ prevents the use of standard techniques unless one constructs
combinations of the characteristics of spouses to reduce the size of the
affinity matrix. A possibility to reduce the dimensionality of the model
would be to group castes into larger classes. Since castes are organized
hierarchically, the analyst could for instance define a threshold, grouping
castes ranked below the threshold into one class and the others into a
second class. The associated affinity matrix would have size $12\times 12$,
resulting in $144$ parameters to be estimated with 284 couples, that is a
ratio of roughly 1 parameter for 2 observations. However, this aggregation
of castes into larger classes would impose restrictions on the role of
castes in matching affinity. The chosen aggregation implies indeed that a
man of say caste 1 has the same affinity for a woman of any caste within the
same class. Given the rarity of inter-caste marriages and the evidence of
same-caste preferences documented in Banerjee et al. (2013), this assumption
does not seem to be justified. As an alternative, the low-rank optimal
transport method introduced in this paper allows one to perform the
estimation of the affinity matrix while selecting only the relevant
combinations of characteristics that matter for the matching affinities.  }

{\normalsize Figure \ref{IndiaCV} indicates that $\lambda =0.2$ gives
slightly lower values of the CV errors of both $\mathcal{W}\left( A\right) -%
\mathbb{E}_{\hat{\pi}}\left[ \Phi _{A}\left( X,Y\right) \right] $ and $\Vert
\mathbb{E}_{\pi ^{A}}\left[ XY^{\top }\right] -\mathbb{E}_{\hat{\pi}}\left[
XY^{\top }\right] \Vert _{F}$ than $\lambda =0.1$ or $\lambda =0.3$. We
therefore select $\lambda =0.2$ for the analysis. The resulting affinity
matrix has rank 10 as indicated in Figure \ref{IndiaSingular} such that out
of 19 possible dimensions of interaction only 10 are relevant. The first
three dimensions together account for about 50\% of the total matching
affinity as shown in Figure \ref{IndiaSingular}.  }

{\normalsize Inspection of the loadings of the first three dimensions,
reported in Table \ref{IndiaTable}, clearly reveals the importance of castes
in matching affinities. The first 3 dimensions on both sides have indeed
large loadings on castes dummies. In the first dimension for instance, the
dummy variable for being of the \textquotedblleft Brahmin\textquotedblright\
caste has by far the largest loading, i.e. approximately 0.80, for both men
and women. There is therefore a considerable loss in matching affinity for
someone of the \textquotedblleft Brahmin\textquotedblright\ caste to marry
outside of his/her caste. Although less pronounced, we also find evidence
for same-caste affinity for the \textquotedblleft Kayastha,"
\textquotedblleft Baisya," and \textquotedblleft Sagdope\textquotedblright\
castes\footnote{%
The other four main castes only represent a very small fraction of the
sample, less than 10\% together, which probably partly explains why we do
not find significant results for these castes.}, as indicated by the
relatively large loadings with same sign for men and women on the second
dimension (for all three castes) and on the third (for the latter). These
results tend to corroborate Banerjee et al. (2013)'s finding of same-caste
marriage preferences. However, our results also reveal two important new
findings. First, the \textquotedblleft Baisya\textquotedblright\ and
\textquotedblleft Sagdope\textquotedblright\ castes both have positive
loadings on the second dimension, which suggests a significant inter-caste
matching affinity between spouses of these two castes. Second, in contrast,
the \textquotedblleft Kayastha\textquotedblright\ caste has a negative
loading on the second dimension for both men and women. This indicates a
negative (repulsive) matching affinity between men and women of the
\textquotedblleft Baisya\textquotedblright\ and \textquotedblleft
Sagdope\textquotedblright\ on the one hand and men and women of the
\textquotedblleft Kayastha\textquotedblright\ on the other hand.  }

{\normalsize Interestingly, education does not seem to play as an important
role as in the previous application. However, as noted in Banerjee et al.
(2013), this is probably due to the fact that the sample is representative,
not of the whole population but rather of the Bengali middle-class that
exhibits little variation in educational achievement: 85 percent of men and
women in the sample have indeed at least a bachelor's degree.  }

\section{\protect\normalsize Conclusion and Future Research}

{\normalsize In this paper, we have demonstrated the effectiveness of
rank-constrained estimation techniques when solving inverse optimal
transport problems. Inverse optimal transport problems are often faced with
large dimensionality of the data sets; hence it is crucial to develop
dimensionality reduction techniques. We plan to investigate further
applications of this methodology, including explaining the intensity of
mercantile exchanges between countries by the similarities in their
characteristics, predicting stable matches in online dating platforms, or
understanding the determinants of workers' productivity on the labor market.
We also plan to consider an extension of the present methodology to
nonbipartite networks, which will allow to estimate the transport costs in
minimum cost flow problems, with applications to analyzing urban
transportation demand, as well as link formation in social networks.  }

{\normalsize 
%
%
%
%
}

{\normalsize \newpage{}  }

{\normalsize
\begin{figure}[H]
{\normalsize
\begin{centering}
\includegraphics[scale=0.5]{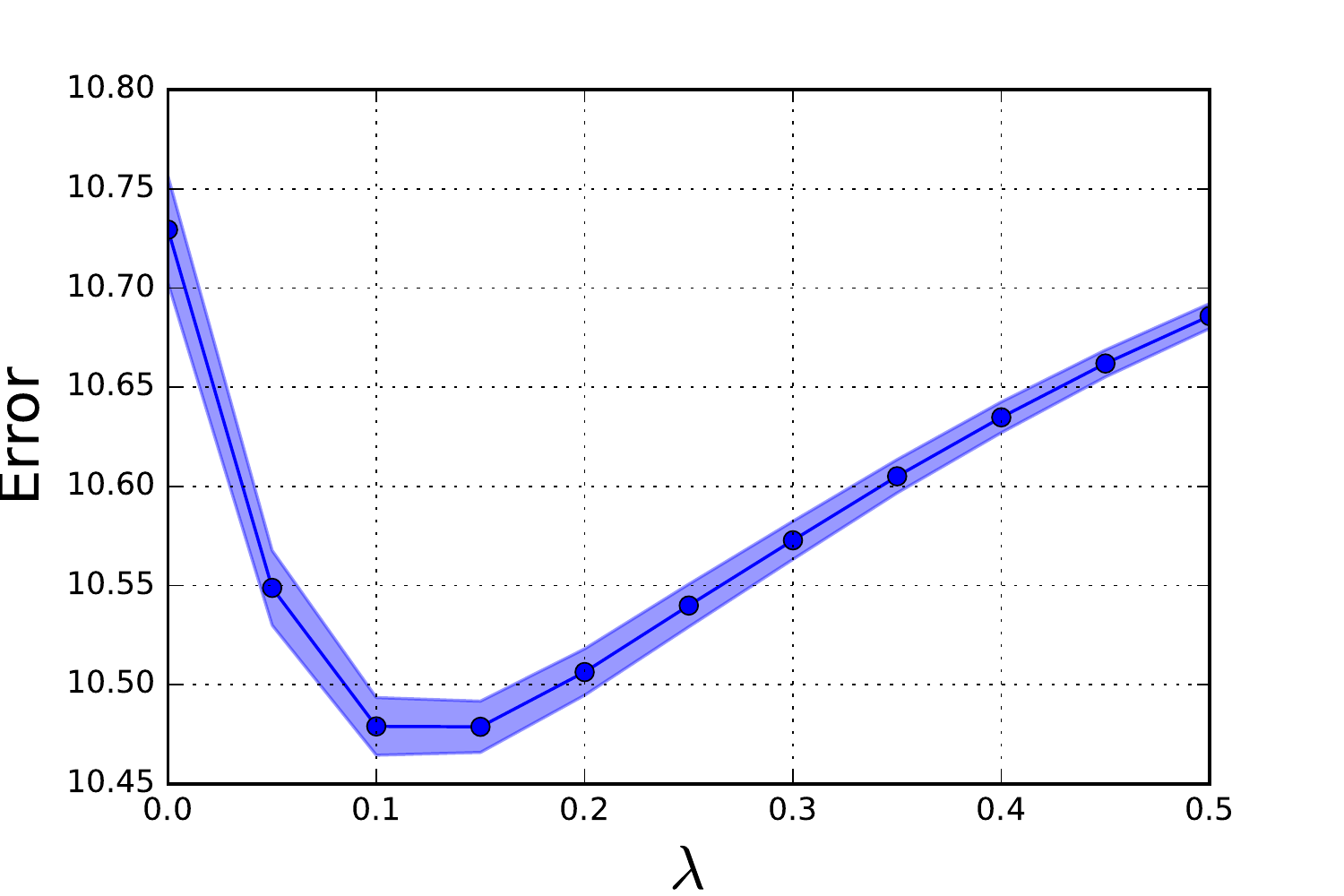}
\includegraphics[scale=0.5]{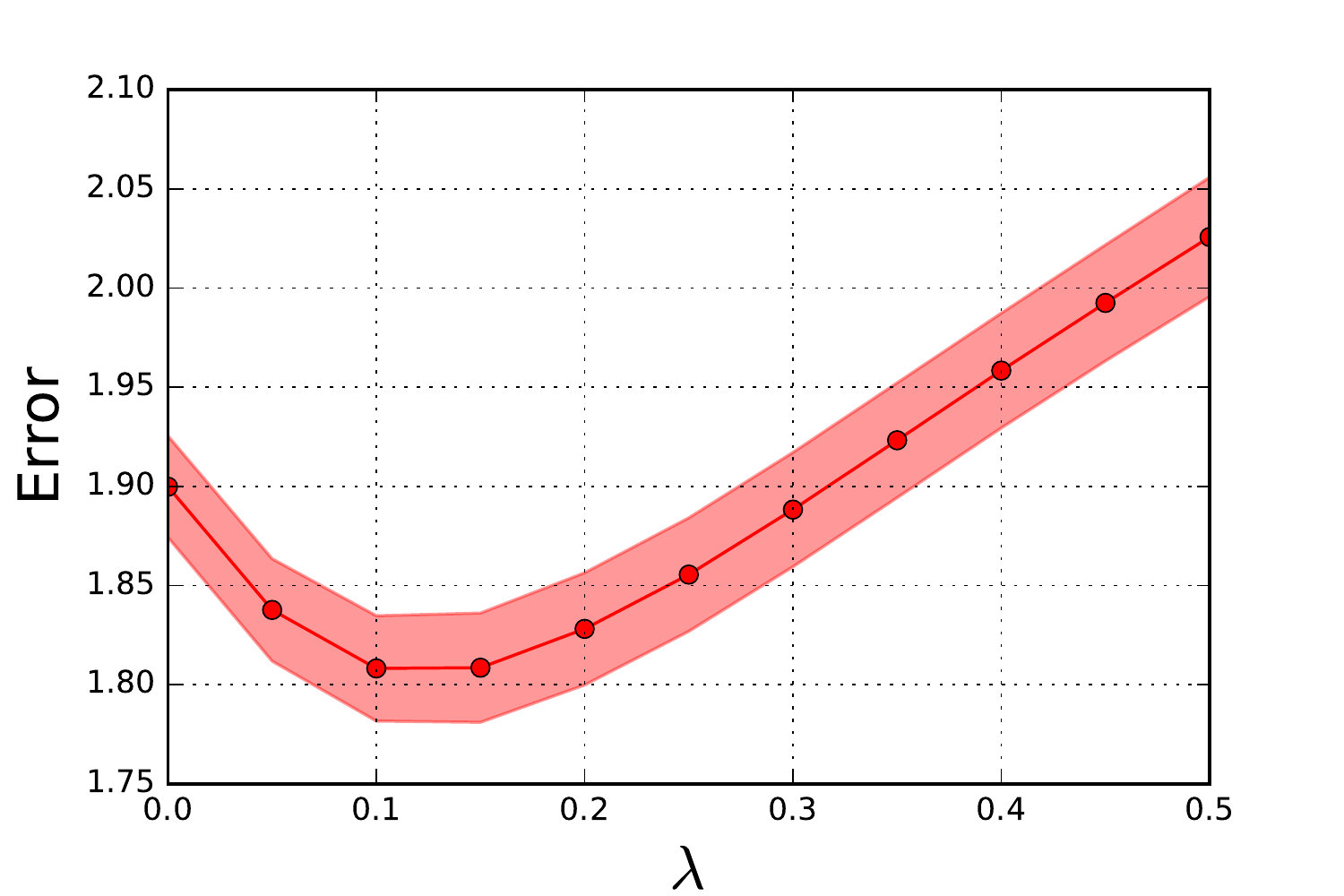}
\par\end{centering}
}
\caption{Cross-validation errors of (left) the negative log-likelihood $%
\mathcal{W}\left(A\right)-\mathbb{E}_{\hat{\protect\pi}}\left[%
\Phi_{A}\left(X,Y\right)\right]$ (right) the covariance mismatch $\Vert%
\mathbb{E}_{\protect\pi^{A}}\left[XY^{\top}\right]-\mathbb{E}_{\hat{\protect%
\pi}}\left[XY^{\top}\right]\Vert_{F}$ .}
\label{DutchCV}
\end{figure}
}

{\normalsize \newpage  }

{\normalsize
\begin{figure}[H]
{\normalsize
\begin{centering}
\includegraphics[scale=0.5]{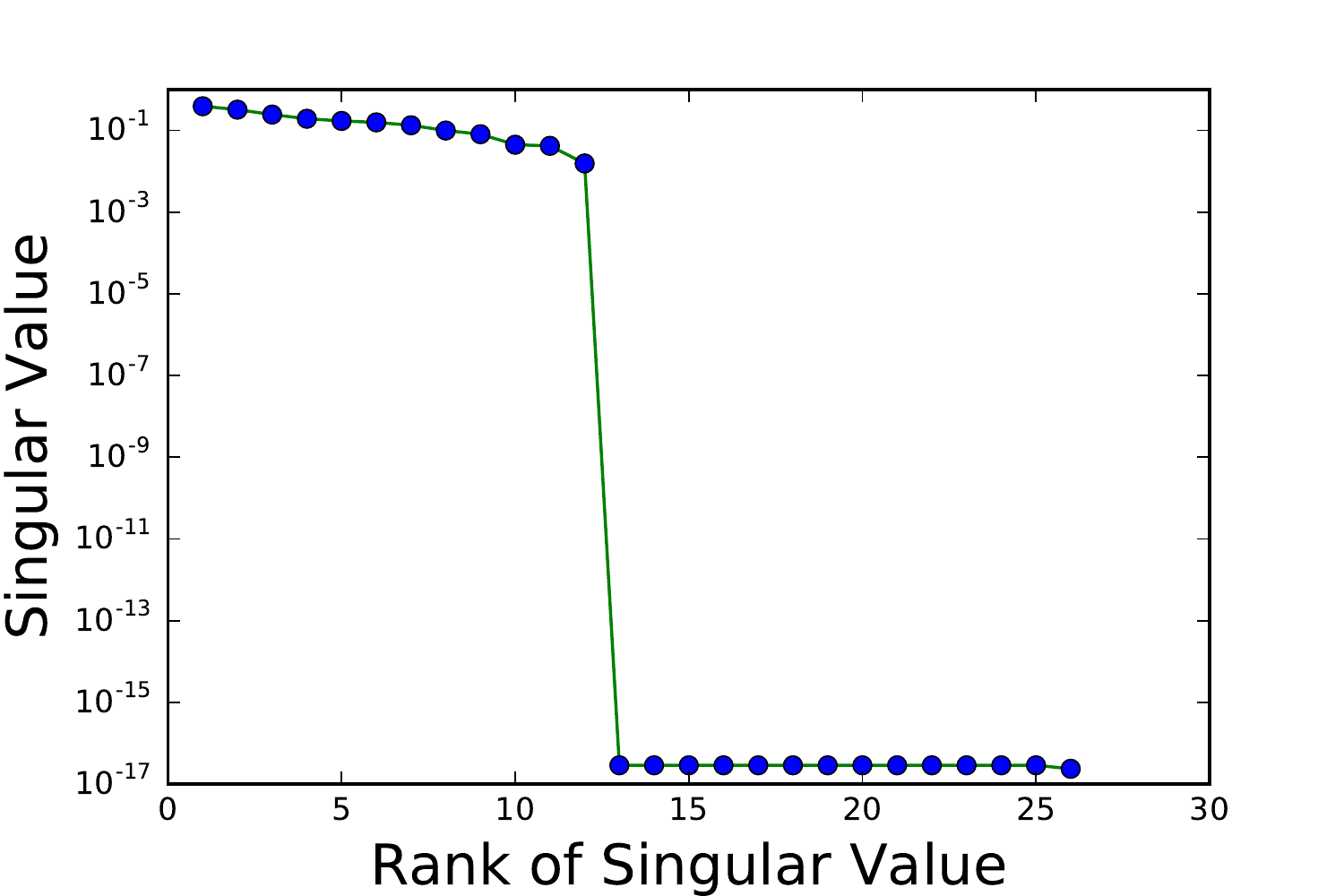}\includegraphics[scale=0.5]{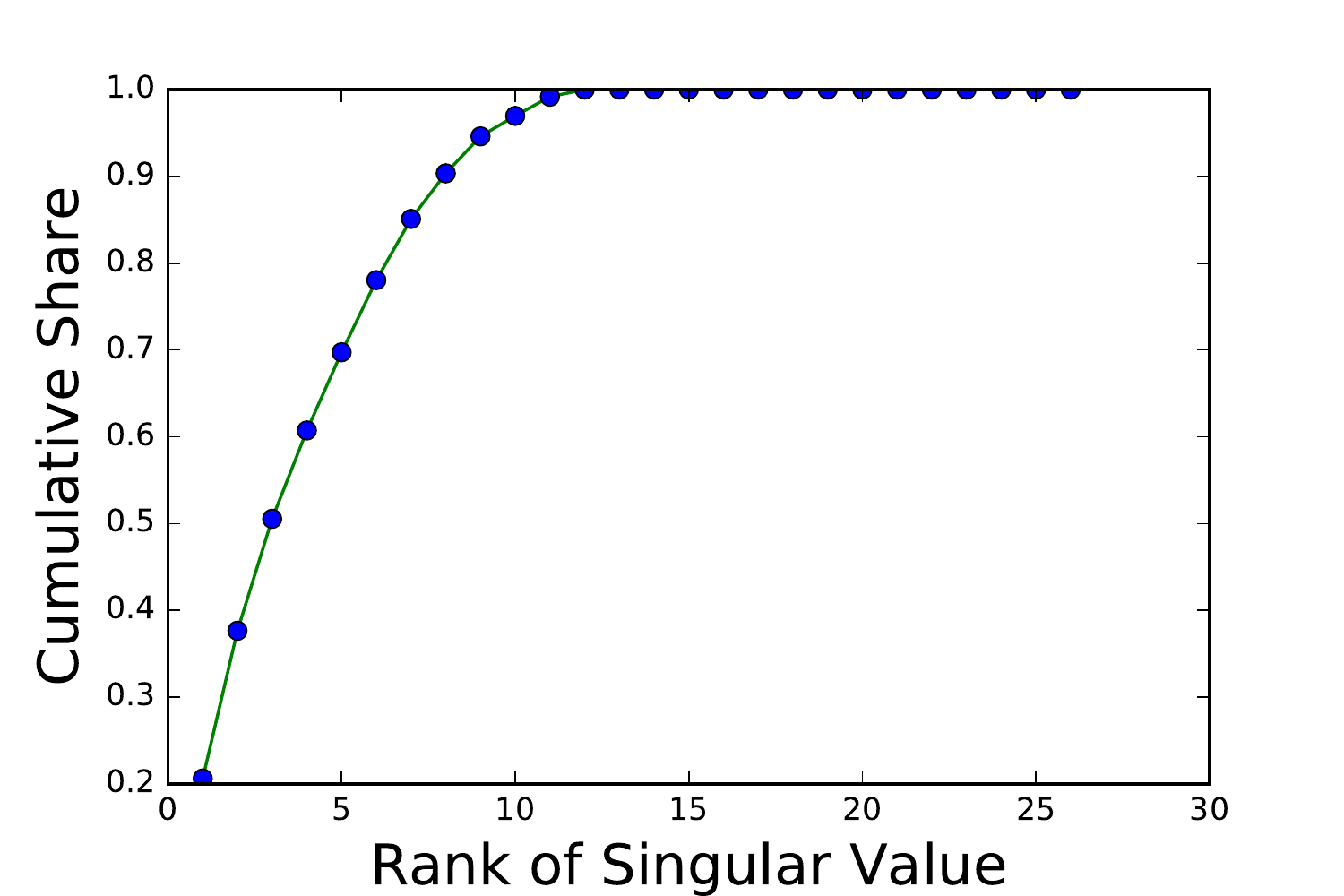}
\par\end{centering}
}
\caption{(Left) singular values (right) cumulative shares.}
\label{DutchSingular}
\end{figure}
}

{\normalsize \newpage  \thispagestyle{empty}
\begin{table}[H]
{\normalsize
\begin{centering}
\begin{tabular}{|c||}
\hline
Singular value\tabularnewline
\hline
\hline
Singular vector\tabularnewline
\hline
Oriented toward people\tabularnewline
\hline
Quick thinker\tabularnewline
\hline
Not easily worried\tabularnewline
\hline
Stubborn, persistent\tabularnewline
\hline
Vivid, vivacious\tabularnewline
\hline
Meticulous\tabularnewline
\hline
Dominant\tabularnewline
\hline
Easily hurt, offended\tabularnewline
\hline
Suspicious\tabularnewline
\hline
Dreamer\tabularnewline
\hline
Diplomatic, tactful\tabularnewline
\hline
Doubts about myself\tabularnewline
\hline
Open to changes\tabularnewline
\hline
Independent, self-reliant\tabularnewline
\hline
Disciplined\tabularnewline
\hline
Irritable, quick tempered\tabularnewline
\hline
Ready to take risk for high possible returns\tabularnewline
\hline
Investments in shares are too risky\tabularnewline
\hline
Ready to borrow money for risky investment\tabularnewline
\hline
Want to be certain my investments are safe\tabularnewline
\hline
Should take greater financial risks\tabularnewline
\hline
Ready to risk losing money to gain money\tabularnewline
\hline
Educational level\tabularnewline
\hline
Height\tabularnewline
\hline
BMI\tabularnewline
\hline
Subjective health\tabularnewline
\hline
\end{tabular}%
\begin{tabular}{|c|c|}
\hline
\multicolumn{2}{|c|}{$s_{1}=0.39$}\tabularnewline
\hline
\hline
$U_{1}$ & $V_{1}$\tabularnewline
\hline
-0.07 & -0.04\tabularnewline
\hline
0.08 & -0.01\tabularnewline
\hline
-0.10 & 0.00\tabularnewline
\hline
0.06 & 0.03\tabularnewline
\hline
0.00 & 0.02\tabularnewline
\hline
-0.12 & -0.05\tabularnewline
\hline
0.05 & 0.06\tabularnewline
\hline
-0.06 & -0.03\tabularnewline
\hline
0.08 & 0.01\tabularnewline
\hline
-0.04 & 0.02\tabularnewline
\hline
-0.06 & 0.07\tabularnewline
\hline
0.06 & 0.13\tabularnewline
\hline
0.10 & 0.03\tabularnewline
\hline
-0.10 & -0.11\tabularnewline
\hline
0.01 & 0.01\tabularnewline
\hline
0.00 & -0.16\tabularnewline
\hline
0.17 & 0.24\tabularnewline
\hline
-0.31 & -0.29\tabularnewline
\hline
0.12 & 0.13\tabularnewline
\hline
0.01 & 0.05\tabularnewline
\hline
-0.06 & -0.07\tabularnewline
\hline
0.10 & 0.09\tabularnewline
\hline
0.85 & 0.83\tabularnewline
\hline
0.06 & 0.08\tabularnewline
\hline
-0.20 & -0.24\tabularnewline
\hline
-0.01 & 0.01\tabularnewline
\hline
\end{tabular}%
\begin{tabular}{|c|c|}
\hline
\multicolumn{2}{|c|}{$s_{2}=0.32$}\tabularnewline
\hline
\hline
$U_{2}$ & $V_{2}$\tabularnewline
\hline
-0.08 & 0.19\tabularnewline
\hline
0.25 & -0.12\tabularnewline
\hline
0.29 & 0.11\tabularnewline
\hline
0.11 & 0.14\tabularnewline
\hline
0.04 & 0.19\tabularnewline
\hline
-0.01 & 0.04\tabularnewline
\hline
-0.09 & -0.21\tabularnewline
\hline
0.42 & 0.71\tabularnewline
\hline
0.16 & 0.14\tabularnewline
\hline
-0.08 & 0.23\tabularnewline
\hline
0.07 & -0.10\tabularnewline
\hline
-0.31 & -0.35\tabularnewline
\hline
0.12 & 0.04\tabularnewline
\hline
0.31 & 0.04\tabularnewline
\hline
0.52 & 0.17\tabularnewline
\hline
0.08 & -0.17\tabularnewline
\hline
-0.02 & -0.07\tabularnewline
\hline
0.05 & -0.07\tabularnewline
\hline
-0.16 & -0.03\tabularnewline
\hline
-0.12 & 0.02\tabularnewline
\hline
-0.03 & -0.05\tabularnewline
\hline
0.06 & -0.02\tabularnewline
\hline
0.12 & 0.12\tabularnewline
\hline
0.12 & 0.01\tabularnewline
\hline
0.17 & 0.18\tabularnewline
\hline
-0.15 & -0.06\tabularnewline
\hline
\end{tabular}%
\begin{tabular}{|c|c|}
\hline
\multicolumn{2}{|c|}{$s_{3}=0.24$}\tabularnewline
\hline
\hline
$U_{3}$ & $V_{3}$\tabularnewline
\hline
0.04 & -0.08\tabularnewline
\hline
-0.04 & 0.05\tabularnewline
\hline
0.08 & -0.07\tabularnewline
\hline
-0.11 & 0.02\tabularnewline
\hline
-0.23 & 0.11\tabularnewline
\hline
0.16 & 0.17\tabularnewline
\hline
0.00 & 0.08\tabularnewline
\hline
0.02 & -0.02\tabularnewline
\hline
0.11 & 0.04\tabularnewline
\hline
0.00 & 0.04\tabularnewline
\hline
0.00 & -0.03\tabularnewline
\hline
0.17 & 0.09\tabularnewline
\hline
-0.22 & 0.05\tabularnewline
\hline
0.01 & -0.09\tabularnewline
\hline
0.01 & -0.11\tabularnewline
\hline
-0.02 & -0.03\tabularnewline
\hline
0.27 & 0.29\tabularnewline
\hline
0.48 & 0.53\tabularnewline
\hline
-0.17 & -0.08\tabularnewline
\hline
0.42 & 0.24\tabularnewline
\hline
-0.11 & -0.18\tabularnewline
\hline
-0.38 & -0.48\tabularnewline
\hline
0.27 & 0.21\tabularnewline
\hline
-0.12 & -0.25\tabularnewline
\hline
0.16 & 0.29\tabularnewline
\hline
-0.21 & -0.17\tabularnewline
\hline
\end{tabular}
\par\end{centering}
}
\caption{Loadings of the top three relevant dimensions of matching
affinities, Dutch couples.}
\label{DutchTable}
\end{table}
}

{\normalsize \newpage  }

{\normalsize
\begin{figure}[H]
{\normalsize
\begin{centering}
\includegraphics[scale=0.5]{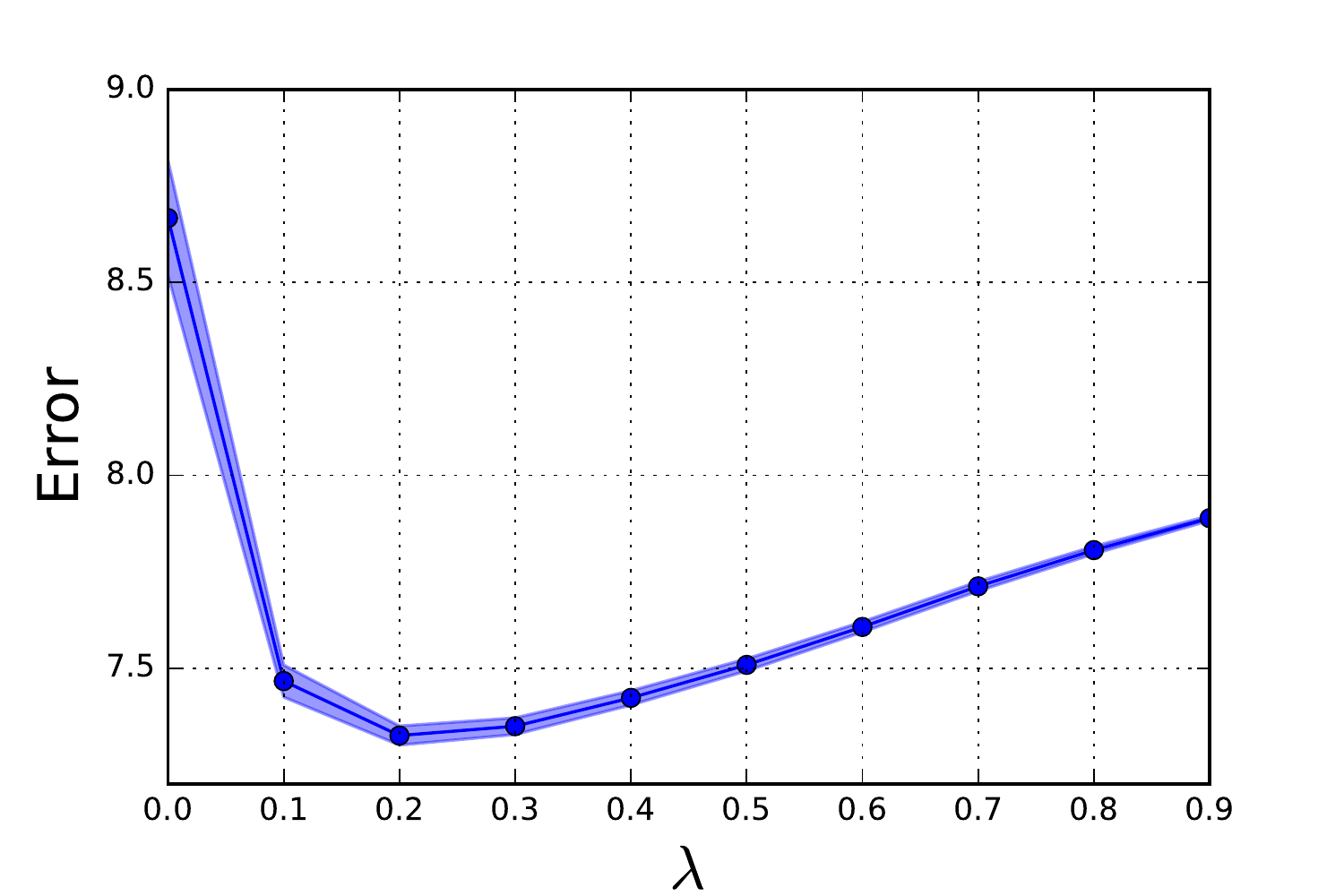}\includegraphics[scale=0.5]{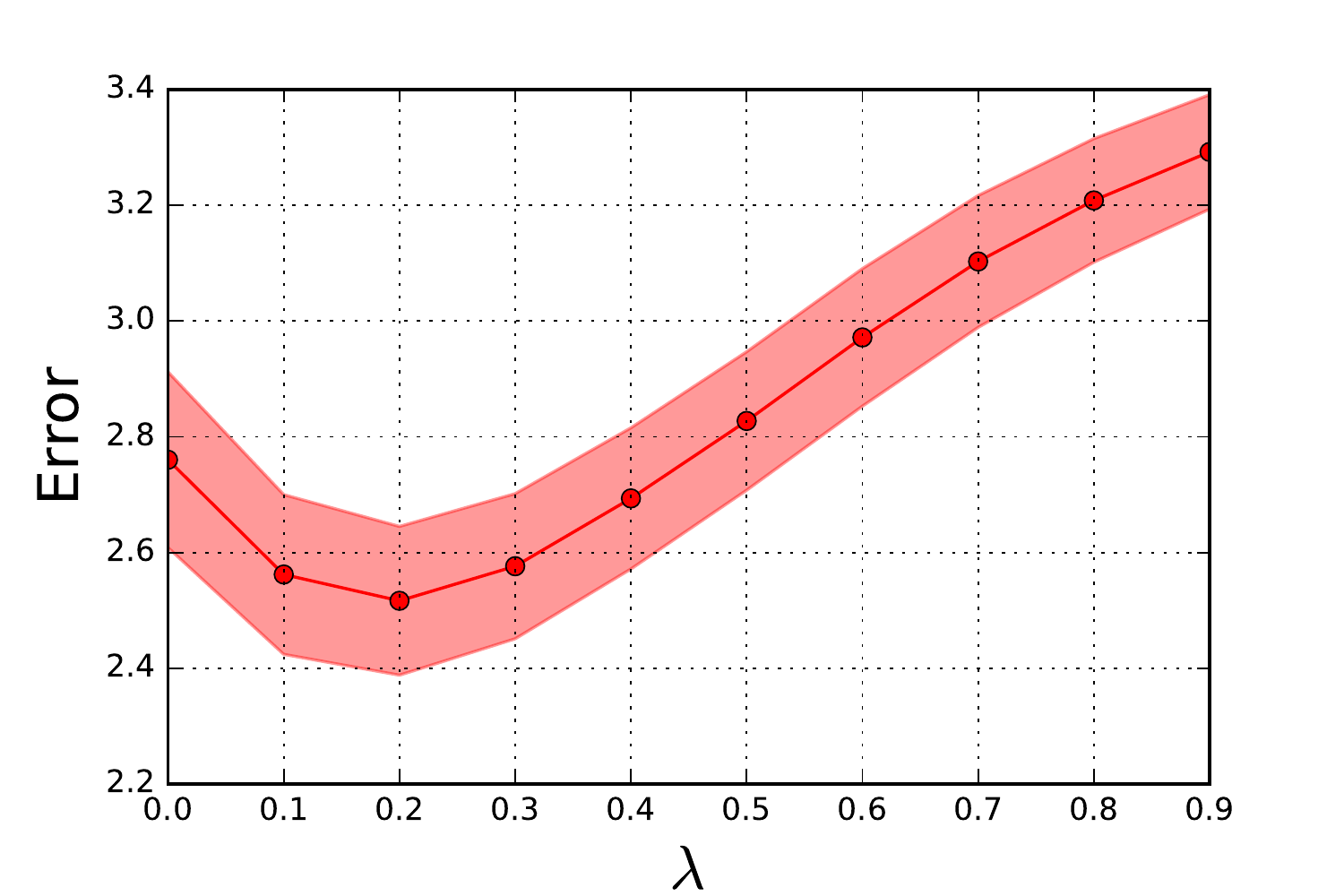}
\par\end{centering}
}
\caption{Cross-validation errors of (left) the negative log-likelihood $%
\mathcal{W}\left(A\right)-\mathbb{E}_{\hat{\protect\pi}}\left[%
\Phi_{A}\left(X,Y\right)\right]$ (right) the covariance mismatch $\Vert%
\mathbb{E}_{\protect\pi^{A}}\left[XY^{\top}\right]-\mathbb{E}_{\hat{\protect%
\pi}}\left[XY^{\top}\right]\Vert_{F}$ .}
\label{IndiaCV}
\end{figure}
}

{\normalsize \newpage  }

{\normalsize
\begin{figure}[H]
{\normalsize
\begin{centering}
\includegraphics[scale=0.5]{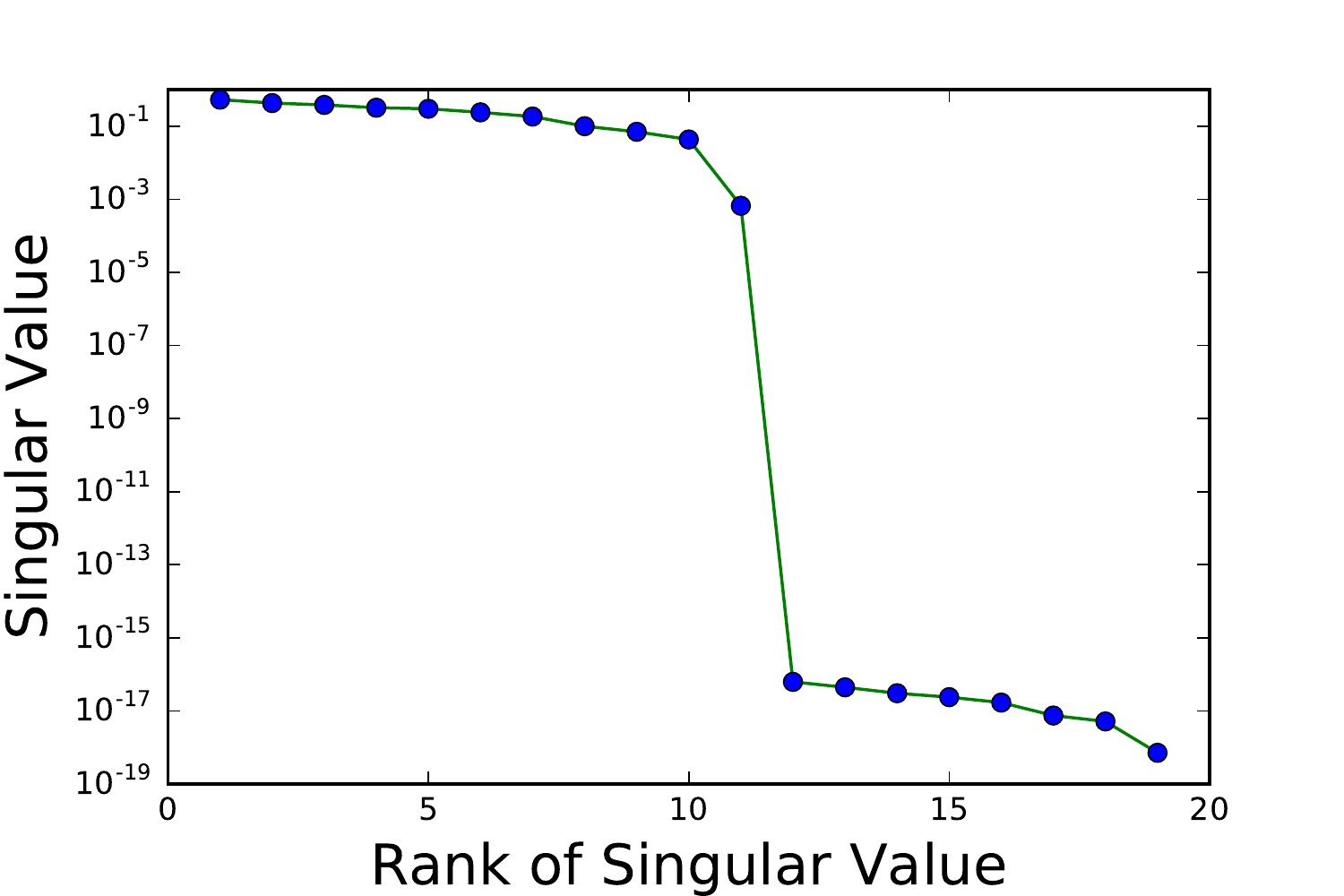}\includegraphics[scale=0.5]{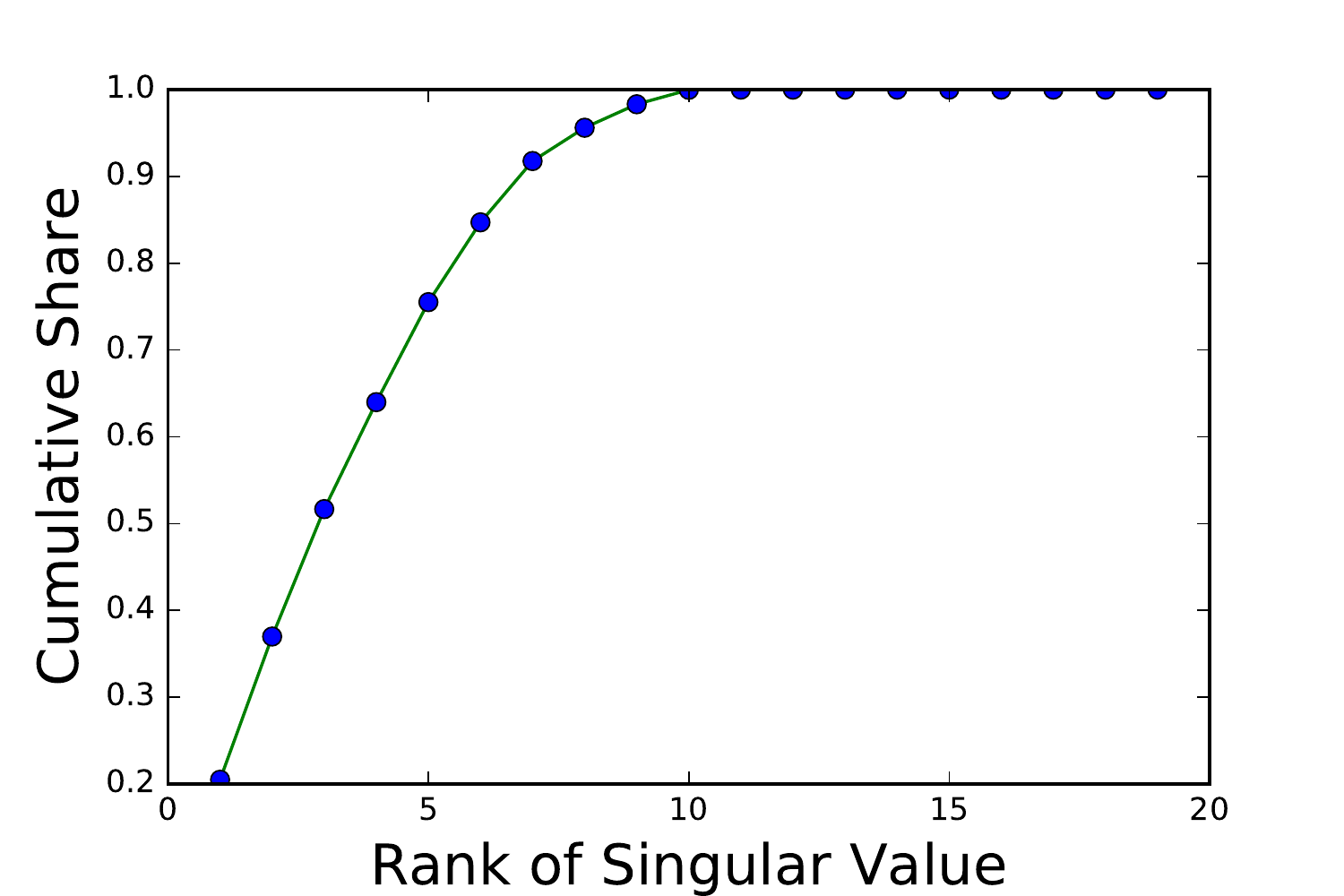}
\par\end{centering}
}
\caption{(Left) singular values (right) cumulative shares.}
\label{IndiaSingular}
\end{figure}
}

{\normalsize \newpage
\begin{table}[H]
{\normalsize
\begin{centering}
\begin{tabular}{|c||}
\hline
Singular value\tabularnewline
\hline
\hline
Singular vector\tabularnewline
\hline
Height\tabularnewline
\hline
Height missing\tabularnewline
\hline
Educational level\tabularnewline
\hline
Family from west Bengal\tabularnewline
\hline
Number of older brothers\tabularnewline
\hline
Number of younger brothers\tabularnewline
\hline
Number of older sisters\tabularnewline
\hline
Number of younger sisters\tabularnewline
\hline
Income\tabularnewline
\hline
Per capita consumption\tabularnewline
\hline
Per capita consumption missing\tabularnewline
\hline
Castes: 1 Brahmin\tabularnewline
\hline
2 Baidya\tabularnewline
\hline
3 Kshatriya\tabularnewline
\hline
4 Kayastha\tabularnewline
\hline
5 Baisya\tabularnewline
\hline
6 Sagdope\tabularnewline
\hline
7 Others\tabularnewline
\hline
8 Scheduled castes\tabularnewline
\hline
\end{tabular}%
\begin{tabular}{|c|c|}
\hline
\multicolumn{2}{|c|}{$s_{1}=0.53$}\tabularnewline
\hline
\hline
$U_{1}$ & $V_{1}$\tabularnewline
\hline
-0.22 & -0.13\tabularnewline
\hline
-0.02 & 0.27\tabularnewline
\hline
0.09 & 0.01\tabularnewline
\hline
0.12 & 0.13\tabularnewline
\hline
0.08 & -0.05\tabularnewline
\hline
0.03 & -0.03\tabularnewline
\hline
-0.09 & -0.08\tabularnewline
\hline
-0.07 & 0.04\tabularnewline
\hline
-0.11 & 0.06\tabularnewline
\hline
0.02 & -0.01\tabularnewline
\hline
0.06 & -0.04\tabularnewline
\hline
0.80 & 0.79\tabularnewline
\hline
-0.10 & -0.03\tabularnewline
\hline
-0.03 & 0.05\tabularnewline
\hline
-0.30 & -0.38\tabularnewline
\hline
-0.34 & -0.22\tabularnewline
\hline
-0.06 & -0.09\tabularnewline
\hline
-0.13 & -0.10\tabularnewline
\hline
-0.10 & -0.20\tabularnewline
\hline
\end{tabular}%
\begin{tabular}{|c|c|}
\hline
\multicolumn{2}{|c|}{$s_{2}=0.43$}\tabularnewline
\hline
\hline
$U_{2}$ & $V_{2}$\tabularnewline
\hline
-0.05 & 0.13\tabularnewline
\hline
-0.01 & 0.12\tabularnewline
\hline
0.26 & 0.30\tabularnewline
\hline
0.16 & 0.32\tabularnewline
\hline
-0.06 & -0.17\tabularnewline
\hline
-0.01 & -0.16\tabularnewline
\hline
-0.06 & 0.13\tabularnewline
\hline
-0.19 & -0.09\tabularnewline
\hline
0.24 & -0.02\tabularnewline
\hline
0.06 & 0.13\tabularnewline
\hline
-0.06 & 0.22\tabularnewline
\hline
-0.07 & -0.15\tabularnewline
\hline
-0.13 & 0.01\tabularnewline
\hline
-0.05 & 0.10\tabularnewline
\hline
-0.50 & -0.41\tabularnewline
\hline
0.40 & 0.39\tabularnewline
\hline
0.57 & 0.48\tabularnewline
\hline
-0.19 & -0.23\tabularnewline
\hline
0.06 & -0.06\tabularnewline
\hline
\end{tabular}%
\begin{tabular}{|c|c|}
\hline
\multicolumn{2}{|c|}{$s_{3}=0.38$}\tabularnewline
\hline
\hline
$U_{3}$ & $V_{3}$\tabularnewline
\hline
-0.21 & -0.20\tabularnewline
\hline
0.07 & 0.00\tabularnewline
\hline
-0.27 & -0.08\tabularnewline
\hline
-0.37 & -0.42\tabularnewline
\hline
0.08 & -0.08\tabularnewline
\hline
0.03 & 0.18\tabularnewline
\hline
0.05 & 0.10\tabularnewline
\hline
0.17 & -0.01\tabularnewline
\hline
-0.21 & -0.17\tabularnewline
\hline
-0.25 & -0.03\tabularnewline
\hline
0.00 & -0.10\tabularnewline
\hline
0.12 & 0.09\tabularnewline
\hline
-0.34 & -0.37\tabularnewline
\hline
-0.11 & -0.05\tabularnewline
\hline
-0.28 & -0.34\tabularnewline
\hline
0.51 & 0.57\tabularnewline
\hline
-0.25 & -0.20\tabularnewline
\hline
0.15 & 0.11\tabularnewline
\hline
0.21 & 0.22\tabularnewline
\hline
\end{tabular}
\par\end{centering}
}
\caption{Loadings of the top three relevant dimensions of matching
affinities, Indian couples.}
\label{IndiaTable}
\end{table}
}


\begin{thebibliography}{99}
\bibitem{barrio} {\normalsize del Barrio, E., Gin\'{e}, E., and Matr\'{a}n,
C. (1999). Central limit theorems for the Wasserstein distance between the
empirical and the true distributions. \textit{Annals of Probability},
1009--1071.  }

\bibitem{Banerjee13} {\normalsize Banerjee, A., Duflo, E., Ghatak, M., and
Lafortune, J. (2013). Marry for what? Caste and mate selection in modern
India. \textit{American Economic Journal: Microeconomics}, 5(2), 33--72.  }

\bibitem{B} {\normalsize Becker, G. S. (1973). A theory of marriage: Part I.
\textit{The Journal of Political Economy}, 813--846.  }

\bibitem{benamou} {\normalsize Benamou, J. D., Carlier, G., Cuturi, M.,
Nenna, L., and Peyr\'{e}, G. (2015). Iterative Bregman projections for
regularized transportation problems. \textit{SIAM Journal on Scientific
Computing}, 37(2), A1111--A1138.  }

\bibitem{BDM} {\normalsize Burkard, R. E., Dell'Amico, M., and Martello, S.
(2009). \textit{Assignment Problems, Revised Reprint}. SIAM.  }

\bibitem{cattell} {\normalsize Cattell, R. B., Cattell, A. K., and Cattell,
H. E. P. (1993). 16PF fifth edition questionnaire. \textit{Champaign, IL:
Institute for Personality and Ability Testing.}  }

\bibitem{COQ} {\normalsize Chiappori, P. A., Oreffice, S., and
Quintana-Domeque, C. (2012). Fatter attraction: anthropometric and
socioeconomic matching on the marriage market. \textit{Journal of Political
Economy}, 120(4), 659--695.  }

\bibitem{CSW} {\normalsize Chiappori, P. A., Salani\'{e}, B., and Weiss, Y.
(2016). Assortative matching on the marriage market: a structural
investigation. \emph{American Economic Review}, forthcoming.  }

\bibitem{Donkers} {\normalsize Donkers, B. and van Soest, A. (1999),
Subjective measures of household preferences and financial decisions.
\textit{Journal of Economic Psychology}, 20(6), 613--642.  }

\bibitem{DG} {\normalsize Dupuy, A. and Galichon, A. (2014). Personality
traits and the marriage market. \textit{Journal of Political Economy},
122(6), 1271--1319.  }

\bibitem{DupuyGalichonAES} {\normalsize Dupuy, A. and Galichon, A. (2015).
Canonical correlation and assortative matching: A remark. \textit{Annals of
Economics and Statistics}, (119-120), 375--383.  }

\bibitem{fazel} {\normalsize Fazel, M. (2002). \textit{Matrix rank
minimization with applications} (Doctoral dissertation, Stanford
University).  }

\bibitem{Fox10} {\normalsize Fox, J. T. (2010). Identification in matching
games. \textit{Quantitative Economics}, 1(2), 203--254.  }

\bibitem{galichon} {\normalsize Galichon, A. (2016). \textit{Optimal
Transport Methods in Economics}. Princeton University Press.  }

\bibitem{GalichonSalanie} {\normalsize Galichon, A. and Salani\'{e}, B.
(2015). Cupid's invisible hand: Social surplus and identification in
matching models. \textit{Available at SSRN 1804623.}  }

\bibitem{GST} {\normalsize Gautier, P. A., Svarer, M., and Teulings, C. N.
(2010). Marriage and the city: Search frictions and sorting of singles.
\textit{Journal of Urban Economics}, 67(2), 206--218.  }

\bibitem{Jepsen} {\normalsize Jepsen, L. K. (2005). The relationship between
wife's education and husband's earnings: Evidence from 1960 to 2000. \textit{%
Review of Economics of the Household}, 3(2), 197--214.  }

\bibitem{Kalmijn} {\normalsize {Kalmijn, M. (1998). Intermarriage and
homogamy: Causes, patterns, trends. \textit{Annual Review of Sociology},
395--421.} }

\bibitem{LS} {\normalsize Lam, D. and Schoeni, R. F. (1993). Effects of
family background on earnings and returns to schooling: evidence from
Brazil. \textit{Journal of Political Economy}, 710--740. }

\bibitem{LS2} {\normalsize Lam, D. and Schoeni, R. F. (1994). Family ties
and labor markets in the United States and Brazil. \textit{Journal of Human
Resources}, 1235--1258.  }

\bibitem{recht} {\normalsize Recht, B., Fazel, M., and Parrilo, P. A.
(2010). Guaranteed minimum-rank solutions of linear matrix equations via
nuclear norm minimization. \textit{SIAM Review}, 52(3), 471--501.  }

\bibitem{ShapleyShubik} {\normalsize Shapley, L. S. and Shubik, M. (1971).
The assignment game I: The core. \textit{International Journal of Game Theory%
}, 1(1), 111--130.  }

\bibitem{SL} {\normalsize Suen, W. and Lui, H. K. (1999). A direct test of
the efficient marriage market hypothesis. \textit{Economic Inquiry}, 37(1),
29--46. }

\bibitem{Terviö} {\normalsize Tervi\"{o}, M. (2003). \textit{Studies of
Talent Markets} (Doctoral dissertation, MIT). }

\bibitem{prox10} {\normalsize Toh, K. C. and Yun, S. (2010). An accelerated
proximal gradient algorithm for nuclear norm regularized linear least
squares problems. \textit{Pacific Journal of Optimization}, 6(615--640), 15.
}

\bibitem{villani03} {\normalsize Villani, C. (2003). \textit{Topics in
optimal transportation} (No. 58). American Mathematical Society.  }

\bibitem{villani} {\normalsize Villani, C. (2008). \textit{Optimal
transport: old and new} (Vol. 338). Springer Science \& Business Media.  }

\bibitem{watson} {\normalsize Watson, G. A. (1992). Characterization of the
subdifferential of some matrix norms. \textit{Linear algebra and its
applications}, 170, 33--45.  }
\end{thebibliography}
\end{document}